\documentclass[aip,jcp,reprint,final]{revtex4-1}

\usepackage{graphicx}
\usepackage{dcolumn}
\usepackage{bm}

\usepackage{mathrsfs}
\usepackage{amsmath}
\usepackage{amssymb}

\newcommand{\arccosh}{\arccos \! \text{h} }

\begin{document}

\title{The influence of line tension on the formation of liquid bridges}

\author{F. Dutka}
\author{M. Napi\'orkowski}
\affiliation{Instytut Fizyki Teoretycznej, Uniwersytet Warszawski,  00-681 Warszawa, Ho\.za 69, Poland}

\date{\today}

\begin{abstract}
The formation of liquid bridges between a planar and conical substrates is analyzed macroscopically taking into account the line tension. Depending on the value of the line tension coefficient $\tau$ and 
geometric parameters of the system one observes two different scenarios of liquid bridge formation upon changing the fluid state along the bulk liquid-vapor coexistence. For $\tau > \tau^*$ ($\tau^*<0$) 
there is a first-order transition to a state with infinitely thick liquid bridge. For $\tau<\tau^*$ the scenario consists of two steps: first there is a first-order transition to a state with liquid 
bridge of finite thickness which upon further increase of temperature is followed by continuous growth of the thickness of the bridge to infinity. In addition to constructing the relevant phase diagram 
we examine the dependence of the width of the bridge on thermodynamic and geometric parameters of the system. 
\end{abstract}


\maketitle


\section{Introduction} 
In this note we investigate the phase diagram of a fluid enclosed between two infinite walls: one planar and one conical. Such a system resembling the atomic force microscope geometry has been analyzed 
in different contexts \cite{Butt1,Butt3,Jang2}. Particular emphasis has been put on the structure of the  phase diagram which displays a phase characterized by the presence of a liquid bridge formed 
between the walls \cite{Dutka2}. 

The mean curvature of the meniscus of the bridge is given by the Young-Laplace equation \cite{Rowlinson2} and its width is a function of the undersaturation \cite{Jang2,Jang1}. The presence of the bridge 
induces force acting between opposite walls which  can be measured using atomic force microscope \cite{Butt2,Butt3}. It turns out, however, that the line tension can have qualitative influence on the 
phase behavior of such system. This issue has not received much attention in the literature and we discuss it in this note. In particular, we focus on bridge formation and filling transitions along 
the bulk liquid-vapor coexistence where presence of line tension leads to effects similar to what - in a different context - is termed a frustrated complete wetting \cite{[][{ and references therein.}]Bonn1}. 

In the following section we recall the form of the free energy functional of the shape of liquid bridge. This functional and the corresponding equation for equilibrium liquid-vapor interfacial shape 
supplemented by the boundary conditions form the basis of our approach. Their analysis along the bulk liquid-vapor coexistence for different values of the line tension coefficient leads to different 
transition scenarios presented in Section \ref{sec_diagram}. In the last section we summarize our results and point at the possibility of indirect measurement of the width of the bridge. The behavior 
of this width  reflects the transition scenario taking place in the system. 


\section{Shape of a liquid bridge}
\begin{figure}[htb]
 \begin{center}
  \includegraphics{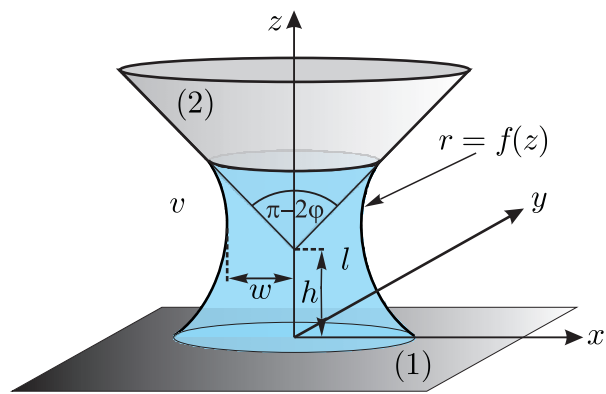}
 \caption{A liquid-like bridge ($l$) surrounded by vapor ($v$) connects the planar $(1)$ and conical $(2)$ walls. 
 \label{bridge} }
 \end{center}
\end{figure}

The system under considerations consists of a fluid confined between two walls, Fig.\,\ref{bridge}. The thermodynamic state of the fluid is located on the bulk liquid-vapor coexistence line. 
The surface of the lower substrate $(1)$ is an infinite plane $z=0$, and the upper substrate $(2)$ is formed by infinite cone whose tip is at distance $h$ from the plane $z=0$. The system is axially 
symmetric around the $z$-axis. In cylindrical coordinates $(r=\sqrt{x^2+y^2}, z \geqslant 0)$ the surface of the cone is described by $r=a(z)=(z-h) \cot \varphi$, where $z \geqslant h$ and 
$\pi-2 \varphi$ is the  opening 
angle of the cone ($0 \leqslant \varphi \leqslant \pi/2$). In our macroscopic approach the grand canonical functional $\Omega([f], T, h, \varphi)$ is a functional of the liquid-vapor interfacial 
shape $r=f(z)$. It is  parametrized  by the temperature $T$ and by geometric parameters $h$ and $\varphi$. Instead of temperature we shall use the angle $\theta$ which fulfills the 
Young equation \cite{Rowlinson2}. The actual contact angles present in the problem fulfill the modified Young equations \cite{Swain1} but it is convenient to use $\Theta$ to reparametrize the 
temperature dependence. We assume that both substrates are made of the same material and are thus characterized by the same angles  $\theta_1 = \theta_2 = \theta$, and the same line tension 
coefficients $\tau_{1lv} = \tau_{2lv} = \tau$. Accordingly, $\cos \theta = (\sigma_{sv} - \sigma_{sl})/\sigma_{lv}$, where $\sigma_{\alpha\beta}$ denote the relevant surface tension coefficients. 
Although for a given system both the line and surface tension coefficients are functions of its thermodynamic state, in our macroscopic analysis they will be varied independently. 
In other words, the parameters $\theta$ and $\tau$ will be considered as independent variables. In particular, the possibility of both signs of $\tau$ will be taken into account. 

At bulk liquid-vapor coexistence the functional of the liquid bridge shape $f(z)$ relative to the state without the bridge $\Delta \Omega[f]$ is given by \cite{Dutka2}: 
\begin{eqnarray}
\label{deltaOmega}
\frac{\Delta \Omega[f]}{2 \pi \sigma_{lv} } =   
     \int dz \, \Theta(f-a) \Big\{ f \sqrt{1+f'^2} \, \Theta(f) \, \Theta(z) \nonumber \\
     -  \cos \theta \Big[\frac{f^2}{2} \, \delta(z) 
     + \frac{a}{\sin \varphi} \, \Theta(z-h) \, \Big] \nonumber \\
      + \tilde \tau  \, \Big[ f \, \delta(z)  +  \cot \varphi \, \Theta(z-h)\, \Big] \Big\}  ,
\end{eqnarray}
where the ratio $\tilde \tau=\tau/\sigma_{lv}$ has dimension of length. The symbols 
$\Theta(z)$ and $\delta(z)$ denote the Heaviside and Dirac function, respectively. 

The equilibrium interfacial shape $\bar f(z)$  minimizes $\Delta \Omega[f]$. This leads to equation
\begin{eqnarray} 
\label{shape}
\frac{1}{\bar f(z) \sqrt{1+\bar f'(z)^2}} - \frac{\text{d}}{\text{d}z} \frac{\bar f'(z)}{\sqrt{1+\bar f'(z)^2}} = 0
\end{eqnarray}  
and two boundary conditions 
\begin{align} \label{boundaries}
\begin{split}
 0 =& \Bigg[ \cos \theta + \frac{\bar f'(z)}{\sqrt{1+\bar f'(z)^2}} - \frac{\tilde \tau}{\bar f(z)}  \Bigg] \Bigg|_{z=0} \, ,  \\
 0 =& \Bigg[\cos \theta - \frac{\sin\varphi+\cos \varphi \bar f'(z)}{\sqrt{1+\bar f'(z)^2}} 
 - \frac{\tilde \tau}{\bar f(z)}\cos \varphi \Bigg] \Bigg|_{z=z_2}  \, .   
\end{split}
\end{align}
The above boundary conditions are equivalent to the modified Young equations for each of the substrates. The coordinate $z_2$ is such that $\bar f(z_2)=a(z_2)$. The lhs of (\ref{shape}) 
is equal to the mean curvature of the interface and  thus the surface of the bridge is a catenoid. The solution of (\ref{shape}) 
\begin{equation}
 \bar f(z) = w \cosh \frac{z-z_1}{w} \, ,
\end{equation}
is parametrized by $z_1$ and $w$, where $w = \bar f(z_1)$ is the minimal value of $\bar f$ which  will be considered to be the width of the bridge. With 
the help of dimensionless quantities $\alpha = z_1 / w$ and $\beta = z_2 / w$ the boundary conditions (\ref{boundaries}) can be rewritten as
\begin{align} 
 \cos \theta = & \, \frac{\tilde \tau}{w \cosh \alpha} + \tanh \alpha \, ,  \label{bound_alpha1} \\
 \cos \theta = & \, \frac{\tilde \tau \cos \varphi}{w \cosh (\beta -\alpha)} + \frac{\sin \varphi}{\cosh (\beta -\alpha)} 
    + \tanh (\beta-\alpha) \cos \varphi \, .  \label{bound_alpha2}
\end{align}
Now the width of the bridge $w$ can be considered to be the following function of $\alpha$ and $\beta$  
parametrized by $h$ and $\varphi$:
\begin{equation} \label{w_alpha}
 w(\alpha, \beta) = \frac{h}{\beta - \tan \varphi \cosh(\beta -\alpha)} \ .
\end{equation}
The relative free energy of the system  is given by 
\begin{equation} \label{omega_alpha}
\Delta \bar \Omega = \pi \sigma_{lv} \, h \, w(\alpha, \beta) \Big[\frac{\tilde \tau}{h} \Big(\cosh \alpha + \cosh (\beta - \alpha)\Big)+1  \Big] \, .
\end{equation} 
Analysis of the above equation enables the construction of the phase diagram; this will be discussed in the next section.


\section{Phase diagram \label{sec_diagram}}
The basis for determining the phase diagram is the knowledge of the equilibrium shapes of the bridge $f=\bar f(z)$ and the corresponding relative grand canonical free energies 
$\Delta \bar \Omega = \Delta \Omega[\bar f]$. Depending on the sign of $\Delta \bar \Omega $ three cases are possible: (a) $\Delta \bar \Omega < 0$ -- the phase with the bridge 
present is favorable, (b) $\Delta \bar \Omega  > 0$ -- phase without bridge is favorable, (c) $\Delta \bar \Omega = 0$ -- the previous phases coexist. 

\begin{figure}[htb]
 \begin{center}
  \includegraphics{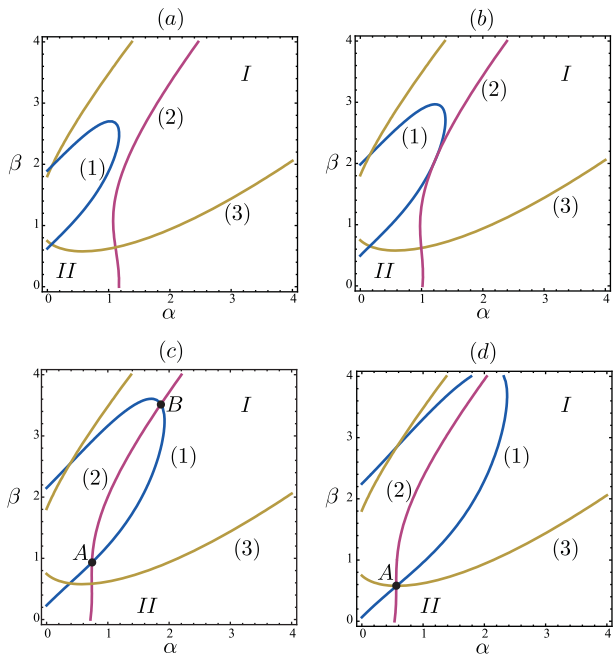}
 \caption{Plots of functions $\beta=\beta(\alpha)$ obtained by numerically solving equations (\ref{bound_alpha1}) and (\ref{bound_alpha2}), curves $(1)$ and $(2)$, respectively. 
Different values of angle $\theta$ are considered: (a) $\theta>\theta^*(\tilde \tau/h, \varphi)$, (b) $\theta=\theta^*(\tilde \tau/h, \varphi)$, (c) 
$\theta^*_0(\varphi) < \theta < \theta^*(\tilde \tau/h, \varphi)$, (d) $\theta = \theta^*_0(\varphi)$ at fixed $\tilde \tau =-h$ and $\varphi = \pi/6$. The curve denoted as 
$(3)$ corresponds to $w(\alpha,\beta)^{-1} = 0$, and divides the  $(\alpha,\beta)$ plane into part corresponding to $w(\alpha,\beta)>0$ (denoted as $I$), and (unphysical) 
part corresponding to $w(\alpha,\beta)<0$ (denoted as $II$). Points of intersection $A$ and $B$ correspond to equilibrium solutions; point $A$ is the solution which 
corresponds to the bridge with a larger width and smaller relative free energy. 
 \label{fil_diag}}
 \end{center}
\end{figure}

The set of equations (\ref{bound_alpha1}) and (\ref{bound_alpha2}) is not solvable analytically. The numerically obtained plots of functions $\beta=\beta(\alpha)$ (parametrized by 
$\theta$, $\tilde \tau$, $h$ and $\varphi$) illustrate the transition scenario taking place at bulk liquid-vapor coexistence, Fig.\,\ref{fil_diag}, for fixed value of the opening 
angle of the cone, $\varphi = \pi/6$, and negative value of the line tension coefficient  $\tilde \tau = -h$. For the angles $\theta>\theta^*(\tilde \tau/h,\varphi)$ the curves 
corresponding to the solutions of equations (\ref{bound_alpha1}) and (\ref{bound_alpha2}) do not intersect, Fig.\,\ref{fil_diag}a. This situation corresponds to the absence of 
the liquid bridge. For $\theta = \theta^*(\tilde \tau/h, \varphi)$, Fig.\,\ref{fil_diag}b, the bridge with a finite width is present, and its relative free energy 
$\Delta \bar \Omega $ is negative. For $\theta^*_0(\varphi)<\theta<\theta^*(\tilde \tau/h,\varphi)$ there are two solutions of (\ref{bound_alpha1}) and (\ref{bound_alpha2}) 
with negative relative free energies corresponding to bridges of different width $w(\alpha,\beta)$. The solution with a larger width is represented by point $A$ on 
Fig.\,\ref{fil_diag}c and has smaller  energy than the one corresponding to point $B$. Upon decreasing the angle $\theta$ towards the angle 
$\theta = \theta^*_0(\varphi)$ the width of the bridge tends to infinity, Fig.\,\ref{fil_diag}d. For $\theta \leqslant \theta^*_0(\varphi)$ the bridge has 
infinite width which corresponds to the whole space between the walls filled with liquid.

\begin{figure}[htb]
 \begin{center}
  \includegraphics{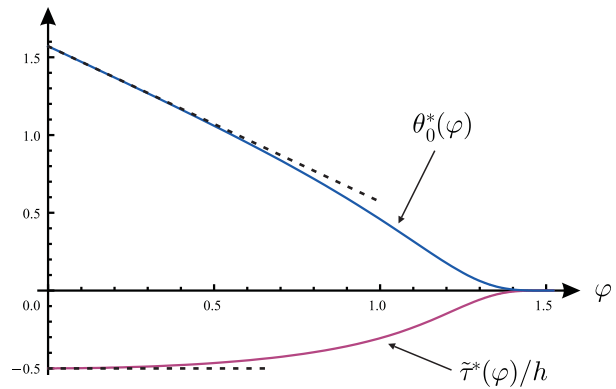}
 \caption{The dependence of the angle $\theta^*_0$ and the dimensionless line tension coefficient $\tilde \tau^*/h$ on the opening angle $\varphi$ (solid lines). For 
$\varphi \ll 1$ the function $\theta^*_0 (\varphi)$ can be approximated by  $\theta^*_0 \simeq \pi/2-\varphi$ and the line tension coefficient by 
$\tilde \tau^*/h \simeq-0.5$ (dashed lines). For $\varphi \to \pi/2$ both functions tend to zero tangentially.
 \label{fig_thetazero}}
 \end{center}
\end{figure}

For particular value of the angle $\theta = \theta^*_0(\varphi)$ the width of the bridge becomes infinite, and equations (\ref{bound_alpha1}) and (\ref{bound_alpha2}) 
do not depend on the line tension coefficient $\tilde \tau$. Their solutions denoted by $\alpha^*$ and $\beta^*$ have the following form 
\begin{align} \label{alpha_star}
 \begin{split}
  \alpha^* = & \, \arccosh \frac{1}{\sin \theta^*_0} \ , \\
  \beta^* = & \, \arccosh \frac{1}{\sin (\theta^*_0 +\varphi)}+\arccosh \frac{1}{\sin \theta^*_0} \, .
\end{split}
\end{align}
After inserting the above expressions to equation $w(\alpha^*,\beta^*)^{-1} = 0$ one gets the equation for the angle $\theta^*_0(\varphi)$
\begin{equation} \label{row_thstar}
\arccosh \frac{1}{\sin (\theta^*_0 +\varphi)}+\arccosh \frac{1}{\sin \theta^*_0} = \frac{\tan \varphi}{\sin(\theta^*_0+\varphi)} \, .
\end{equation}
Its numerical solution is shown on Fig.\,\ref{fig_thetazero}. For $\varphi \ll 1$ the function $\theta^*_0$ can be approximated by 
$\theta^*_0 \simeq \pi/2-\varphi$, and for $\varphi \to \pi/2$ it tends to zero tangentially.

We note that the bridge of finite width can exist for $\theta>\theta^*_0(\varphi)$ provided the free energy of the system (\ref{omega_alpha}) evaluated at 
$\theta=\theta^*_0(\varphi)$ is not greater than zero. This requirement can be rewritten as the condition on the line tension coefficient 
\begin{equation} \label{eq_taustar}
 \tilde \tau \leqslant - h \, \frac{\sin \theta^*_0(\varphi) \sin(\theta^*_0(\varphi) + \varphi)}{\sin \theta^*_0(\varphi) + \sin(\theta^*_0(\varphi) + \varphi)} \equiv \tilde \tau^*(\varphi)\, . 
\end{equation}
Thus for $\tilde \tau \leqslant \tilde \tau^*(\varphi)$ the liquid bridge is present for $\theta \geqslant \theta^*_0(\varphi)$, otherwise there is no bridge in the system. 
For $\varphi \ll 1$ the line tension coefficient can be approximated by the function $\tilde \tau^*/h \simeq-0.5$, and for $\varphi \to \pi/2$ tends to zero tangentially, 
Fig.\,\ref{fig_thetazero}.

In order to find the divergence of the width of the bridge for $\theta \to \theta^*_0(\varphi)$ at fixed $\tilde \tau < \tilde \tau^*(\varphi)$ we expand equations 
(\ref{bound_alpha1}) and (\ref{bound_alpha2}) around $\theta=\theta^*_0(\varphi)$, $\alpha = \alpha^*$ and $\beta=\beta^*$. In the leading order the divergence is given by 
\begin{align}
 w \simeq A(\varphi) \, h \, \frac{\tilde \tau^*(\varphi)-\tilde \tau }{|\tilde \tau^*(\varphi)|} \, \Big[\theta-\theta^*_0(\varphi) \Big]^{-1} \, ,
\end{align}  
where the amplitude 
\begin{equation}
A(\varphi)=\frac{\sin^2(\theta^*_0(\varphi) + \varphi)\,\sin\theta^*_0(\varphi) \,\cos\varphi}{\sin^2(\theta^*_0 + \varphi)+\sin^2\theta^*_0(\varphi)}
\end{equation}
is positive for $0 < \varphi < \pi/2$. 

\begin{figure}[htb]
 \begin{center}
  \includegraphics{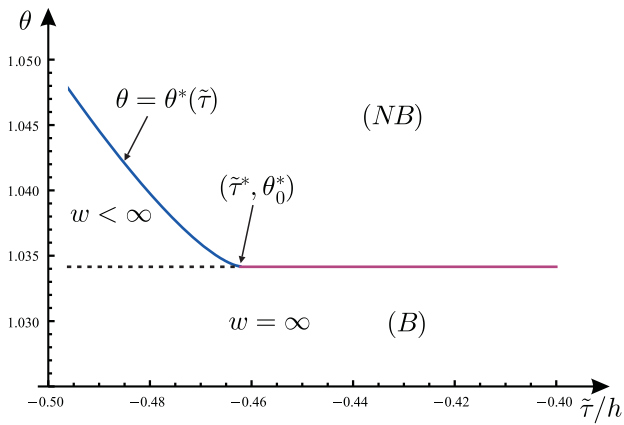}
 \caption{The phase diagram in variables $(\tilde \tau/h, \theta)$ displaying the coexistence line between the phase with no liquid bridge present in the system $(NB)$ 
and with the bridge present in the system $(B)$ (continuous line), $(w<\infty)$ denotes the region in which the width of the bridge is finite, and $(w=\infty)$ -- the region 
where the whole system is filled with liquid. The opening angle is equal $\varphi = \pi/6$, for which $\theta^*_0 \approx 1.034 $ and $\tilde \tau^*/h \approx -0.462$.  
 \label{tau-theta} }
 \end{center}
\end{figure}

To find the angle $\theta = \theta^*(\tilde \tau/h, \varphi)$ at which there is a first order transition from the phase without bridge to phase with the bridge one has to solve 
eqs. (\ref{bound_alpha1}), (\ref{bound_alpha2}), (\ref{w_alpha}) supplemented by the requirement of tangetiality of the curves $\beta(\alpha)$ determined by these equations, Fig.\ref{fil_diag}b. 
The numerically obtained phase diagram, Fig.\,\ref{tau-theta}, displays the
coexistence line between the phase with no bridge in the system $(NB)$ and with the  bridge $(B)$. The coexistence lines for $\tilde \tau \leqslant \tilde \tau^*$ and for 
$\tilde \tau \geqslant \tilde \tau^*$ meet tangentially. For $\theta^*_0 < \theta \leqslant \theta^*$ the width of the bridge is finite $(w<\infty)$ and for 
$\theta \leqslant \theta^*_0$ the whole system is filled with the liquid phase $(w=\infty)$. 

\section{Discussion}
We have shown that a fluid confined between planar and a conical walls may undergo -- upon changing its thermodynamic state  along the bulk liquid-vapor coexistence line -- two different 
transition scenarios leading from the phase without liquid bridge to the phase with liquid bridge of infinite thickness. The type of  scenario depends on the value of dimensionless line 
tension coefficient $\tau/\sigma_{lv}h \equiv \tilde \tau /h$. For $\tilde \tau < \tilde \tau^*(\varphi)$  first the bridge of finite width is formed discontinuously at temperature 
corresponding to angle $\theta=\theta^*(\tilde \tau/h, \varphi)$. Upon further decrease of $\theta$ the width of the bridge increases continuously and at 
$\theta=\theta^*_0(\varphi)$ it becomes infinite; the whole space between substrates is filled with the liquid phase. This scenario is qualitatively similar to the one observed 
experimentally by Takata et al.\cite{Takata1} in a different context. On the other hand, when decreasing the angle $\theta$ at $\tilde \tau > \tilde \tau^*(\varphi)$ one 
observes a discontinuous transition at $\theta=\theta^*_0(\varphi)$. This transition takes the system from the phase without the liquid bridge to the phase with bridge of infinite width. 

To find the width of the bridge one can perform the solvation force measurements using the atomic force microscope \cite{Butt1,Butt3,Jang2} with conical tip. The solvation 
force associated with the liquid bridge $
F = - \partial \Delta \bar \Omega(T,h,\varphi)/ \partial h $ 
can be rewritten in the following $z$-independent form 
\begin{equation}
\label{tt1}
 F = - 2 \pi \sigma_{lg} \, \frac{ \bar{f}(z)^2}{2} \left[\frac{1}{r_1(z)}+\frac{1}{r_2(z)} \right] \, ,
\end{equation}
where the radii of curvature
\begin{align} 
 \begin{split}
  r_1(z) = & \,  \left[ \frac{{\rm d}}{{\rm d}z} \frac{\bar f'(z)}{\sqrt{1+\bar f'(z)^2}} \right]^{-1} \, , \\
  r_2(z) = & \, \bar f(z) \sqrt{1+\bar f'(z)^2} \, ,
 \end{split}
\end{align}
can be evaluated at arbitrary $z \in [0;z_2]$. The $z$-independent expression in (\ref{tt1}) can be presented in a particularly transparent form \cite{Tabrizi1}
\begin{equation}
F = - 2 \pi \sigma_{lg} w \, ,
\end{equation}
where the negative sign indicates that the substrates attract each other once the liquid bridge is formed. Thus the solvation force measurements provide direct information on the width of the liquid bridge. 

\bibliography{bibliography}
\end{document}